\documentclass[a4paper, 10pt, accepted=2021-11-03]{quantumarticle}

\pdfoutput=1

\usepackage[titletoc,title]{appendix}
\usepackage[utf8]{inputenc}
\usepackage[T1]{fontenc}
\usepackage[colorlinks]{hyperref}
\usepackage[english]{babel} 
\usepackage{amsmath}
\usepackage{amsfonts}
\usepackage{amssymb}
\usepackage{amsthm}
\usepackage{euscript}
\usepackage[ruled, linesnumbered, noend]{algorithm2e}
\usepackage[numbers,sort&compress]{natbib}

\newtheorem{theo}{Theorem}

\newtheorem{lemma}{Lemma}

\DeclareMathOperator{\Find}{Find}
\DeclareMathOperator{\Union}{Union}
\DeclareMathOperator{\Grow}{Grow}
\DeclareMathOperator{\Support}{Support}

\newcommand{\e}{\varepsilon}

\title{Almost-linear time decoding algorithm for topological codes}

\makeatletter
\let\@fnsymbol\@arabic
\makeatother

\author{Nicolas Delfosse}
\affiliation{IQIM, California Institute of Technology, Pasadena, CA, USA}
\affiliation{Department of Physics and Astronomy, University of California, Riverside, CA, USA}
\affiliation{Station Q Quantum Architectures and Computation Group, Microsoft Research, Redmond, WA 98052, USA}

\email{nidelfos@microsoft.com}

\author{Naomi H. Nickerson}
\affiliation{Quantum Optics and Laser Science, Blackett Laboratory, Imperial College London, Prince Consort Road, London SW7 2AZ, United Kingdom}

\begin{document}

\begin{abstract}
In order to build a large scale quantum computer, one must be able to  
correct errors extremely fast. We design a fast decoding algorithm 
for topological codes to correct for Pauli errors and erasure
and combination of both errors and erasure. 
Our algorithm has a worst case complexity of $O(n \alpha(n))$, 
where $n$ is the number of physical qubits and 
$\alpha$ is the inverse of Ackermann's function, which is very slowly growing.
For all practical purposes, $\alpha(n) \leq 3$.
We prove that our algorithm performs optimally for errors of weight up to $(d-1)/2$
and for loss of up to $d-1$ qubits, where $d$ is the minimum distance of the code.
Numerically, we obtain a threshold of $9.9\%$ for the 2d-toric code with perfect syndrome measurements and $2.6\%$ with faulty measurements.
\end{abstract}

\maketitle

\section*{Introduction}

\medskip
The main obstacle to the construction of a quantum computer is the unavoidable
presence of errors, which left unchecked quickly destroy quantum information. 
Error correction will therefore be essential to perform meaningful quantum computation.
Experimental efforts have made rapid progress in recent years~\cite{barends2014superconducting,kalb2017entanglement,ballance2014high,debnath2016demonstration,heeres2017implementing,carolan2015universal,hill2015surface,monz2016realization,lekitsch2017blueprint,ibm}, 
and may soon have the capabilities to demonstrate small-scale error correction. 
Topological codes, in particular Kitaev's surface code~\cite{kitaev2003fault}, are currently expected 
to form the core architecture of this first generation of quantum computers, due to their high thresholds 
and their locality. However, to use these codes, we also require a classical decoding algorithm, 
which must process measurement information fast enough to keep pace with the clock-speed
of the quantum device. 
While the question of which codes will be the first to be realized seems to 
be answered, no existing decoder is yet fast enough to match the speeds 
that the first generation of quantum processors will require~\cite{fowler2017:QEC_talk}.

\medskip
Many decoding algorithms that run in polynomial time have developed~\cite{
dennis2002topological,
harrington2004:phd,
dennis2005:phd,
raussenfort2007:ft,
stace2010error,
barrett2010fault,
duclos2010fast,
fowler2012topological,
fowler2012:decoding,
wootton2012high,
duclos2013fault,
duclos2013kitaev,
bravyi2013quantum,
hutter2014:HDRG,
anwar2014fast,
bravyi2014efficient,
hutter2015:improved_HDRG,
wootton2015:simple_decoder,
fowler2015:parallel_MWPM,
watson2015fast,
varsamopoulos2017decoding,
herold2015fault,
torlai2017neural,
tuckett2017ultra,
landahl2011:color_codes_decoding,
wang2009graphical,
sarvepalli2012efficient,
bombin2012:color_codes_decoder,
delfosse2014decoding, 
kubica2018:phd}
but although this is considered efficient, in practice quadratic or cubic complexity
is likely too slow to correct errors faster than they accumulate in a quantum device. 
Minimum weight perfect matching (MWPM) decoder~\cite{dennis2002topological}
is currently the most standard decoder for topological codes, and has a worst case 
complexity of between $O(n^3)$ and $O(n^7)$ depending on the implementation~\cite{kolmogorov2009:blossom}. 
Significant efforts have been made to optimize its performance~\cite{fowler2012topological}, and extend it to more general noise models~\cite{stace2010:loss_long, barrett2010:loss, stace2010:percolation, whiteside2014:loss}. Most notably, Fowler
has achieved large speed improvements~\cite{fowler2015:parallel_MWPM}.
But despite this, further speed-up is required if the decoder is to be practical in a real device. 

\medskip
In this work, we design a decoding algorithm for topological codes 
that runs in the worst case in almost-linear time in the number of physical qubits $n$,
with a high threshold (See Table~\ref{tab:thresholds}).
We focus on the worst case complexity and not only on the average case complexity 
since it is the maximum running time of the decoder that will determine the clock-time 
of the quantum computer.
Our key insight is the use of the Union-Find data-structure 
algorithm~\cite{galler1964:union_find_algorithm, tarjan1975:union_find_complexity} 
that allows us to dynamically keep track of and update the estimation of the error as the decoder runs. 
We obtain a threshold of $9.9\%$ for correction of phase-flip or bit-flip Pauli error 
and $2.6\%$ with faulty syndrome measurement. 
By {\em almost-linear} complexity, we do not mean $O(n \log n)$ but even lower.
Our decoder has a worst-case complexity $O(n \alpha(n))$, where 
$\alpha$ is the inverse of Ackermann's function \cite{tarjan1975:union_find_complexity}. 
Although it is not formally linear, $\alpha$ is so slowly growing that it can be considered 
as a constant. 
If the number of physical qubits used is smaller than the number of atoms in the universe, 
then $\alpha(n) \leq 3$.

 \begin{table}[hb]
\centering
\footnotesize
\caption{Comparison of the Union-Find and MWPM decoders' thresholds under 
phase-flip error.}
\label{tab:thresholds}
\begin{tabular}{c c c}
\hline
 & UF decoder & MWPM decoder\\
\hline
\text{2d-toric code } & $9.9\%$ & $10.3\%$ \cite{dennis2002topological}\\
\hline
\text{2+1d-toric code} & $2.6\%$ & $2.9\%$ \cite{wang2003confinement}\\
\hline
\text{2d-hexagonal color code} & $8.4\%$ & $8.7\%$ \cite{delfosse2014decoding}\\
\hline
\end{tabular}
\end{table}

\medskip
We begin by introducing the surface code in Section~\ref{sec:background}. In Section~\ref{sec:uf} we introduce an outline of the 
decoding algorithm, and give analytic arguments about its performance in Section~\ref{sec:performance}. Sections~\ref{sec:complexity} and~\ref{sec:weighted_growth}
contains our main result, the description of the implementation that can be used to implement our
decoding algorithm in almost-linear time. Finally we present our numerical results, and 
discuss the application of the decoder beyond the surface code in Section~\ref{sec:application}. Further discussion of the complexity scaling,
and numerical simulations is given in the Appendices.

\begin{figure}
\includegraphics[width=0.9\columnwidth]{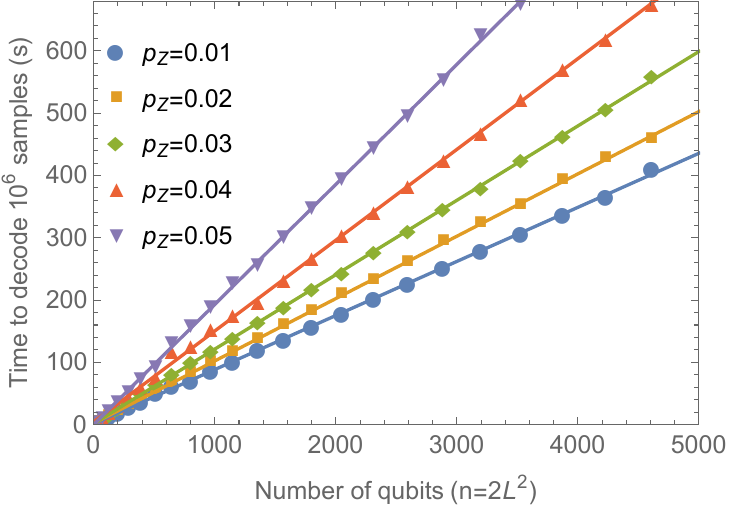}
\caption{\label{fig:time_scaling} {\bf Average running time.} We show the time taken to perform $10^6$ Montecarlo trials of decoding for a 2d toric lattice, under Pauli error and erasure error, with $p_e=0.1$. The average decoder running time increases linearly with the number of qubits. Our decoder was implemented in C, and run on a single 2.9 GHz Intel Core i5 CPU.}
\end{figure}

\section{Background - the surface code \label{sec:background}}

Our decoding algorithm generalizes to any surface code with any genus, 
with or without boundaries~\cite{freedman2001:planar, bravyi1998:planar}
as well as to color codes \cite{bombin2006:color_codes}. For simplicity, we choose to
describe the implementation for the surface code without periodic boundary conditions.  

\medskip
{\bf The surface code: }
The surface code, introduced by Kitaev~\cite{kitaev2003fault}, is a topological code, 
defined on a square lattice of the torus, where a qubit is placed on each edge. 
Denote respectively by $V,E,F$ the set of vertices, edges and faces of the lattice.
The code is defined to be the ground space of the Hamiltonian,
$$
H = -\sum_{v \in V} X_v - \sum_{f \in F} Z_f
$$
There is an operator $X_v$ associated with each vertex $v$ of the lattice
and a plaquette operator $Z_f$ associated with each face $f$.
$X_v$ is the product of the Pauli-$X$ matrices acting on the edges incident to $v$,
{\em i.e.} $X_v = \prod_{e \in v} X_e$, and 
$Z_f = \prod_{e \in f} Z_e$ is the product of the Pauli-$Z$ acting on all edge of the face $f$.
The code space is defined as the simultaneous `+1' eigenstate of these operators $X_v$ and $Z_f$. 
These operators, and any product of them, are called the {\em stabilizers} of the code, and form
the stabilizer group, $S$.

\medskip
{\bf Error model: }
For simplicity, we consider only i.i.d. phase-flip errors, where each qubit is subjected 
to a $Z$-error with probability $p_Z$. The $X$-part of a Pauli error 
can be corrected identically. In addition to Pauli errors, a qubit may be {\em erased}, with probability $p_e$. 
We call the set of all erased qubits, {\em the erasure}, $\e$. 
We use the term erasure to describe the error channel through
which a qubit at a {\em known location} is subjected to a Pauli $Z$ error with probability $1/2$.
The terms {\em erasure}, {\em loss} and {\em leakage} are sometimes used interchangably, 
but physically loss and leakage are two separate mechanisms through which a qubit can be erased.
If we detect that a qubit has left the computational subspace (leakage), or that it is physically missing (lost),
it can be reinitialized or replaced, which corresponds to a random Pauli error after measurement of the stabilizers. 
However, it is worth noting that other physical processes could cause an error at a known location, 
and these can be treated identically as an erasure.

\medskip
{\bf Error correction: }
Error correction proceeds by measuring the stabilizer 
operators $X_v$. When an error $E_Z \in \{I, Z\}^{\otimes n}$ has affected the qubits of the code, 
any stabilizer $X_v$ that anticommutes with the error returns a `-1' outcome. The subset of vertices, $v$, with $-1$ measurement outcomes is called the {\em syndrome}, $\sigma$. 
Given a syndrome, $\sigma$, and an erasure, $\e$, the role of the decoder is to find a correction 
operator, $\cal{C}(\sigma, \e)$, such that $\mathcal{C}(\sigma, \e) E_Z \in S$. That is, when the correction operator is applied to the code, the error is corrected up to a stabilizer.

\section{Union-Find decoder for surface codes\label{sec:uf}}

We begin by introducing an outline of the decoding procedure, without the 
details of its implementation that are required to achieve low complexity.

\begin{figure}
\includegraphics[width=\columnwidth]{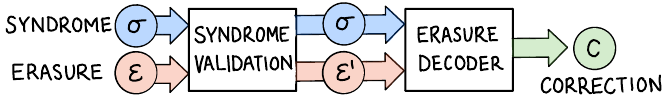}
\caption{{\bf The two stages of decoding.} The erasure decoder proceeds in two stages. We begin with a syndrome, $\sigma$, caused by an error, $E_Z$, that is supported only partially in $\e$. In the first stage, syndrome validation modifies the erasure to $\e'$ in such a way that there is an error $E_Z'$, entirely supported within $\e'$, whose syndrome is also $\sigma$. In the second stage $\sigma$ and $\e'$ are decoded using the erasure decoder. \label{fig:stages}}
\end{figure}

\medskip
Our algorithm is motivated by the fact that erasure errors are much simpler to decode than Pauli errors. 
Simply put, we can describe an erasure error as a Pauli error at a known location. When erasure is the only source of error
the decoding problem is significantly simplified, as errors can only be present within the erasure.
The peeling decoder for erasure introduced in~\cite{delfosse2017:linear_decoding} uses this fact
to find a correction in the case that only erasure errors are present. When Pauli errors can also occur this algorithm will necessarily fail, as the correction now has support outside the erasure. 

\medskip
The decoder is divided into two stages, as shown in Figure~\ref{fig:stages}. 
The goal of the first stage is to take a syndrome generated by both 
Pauli error and erasure, $\e$, and from this generate a modified erasure, $\e'$,
such that there is a valid correction operator supported entirely in $\e'$.
We call this stage {\em syndrome validation}. After the syndrome is validated, 
we can apply the peeling decoder to 
find a correction.

\medskip
To perform syndrome validation, we identify `invalid' clusters of erasures, and iteratively 
grow them until the updated state is correctable by the erasure decoder. 
This idea of growing and merging clusters in order to correct them as locally as 
possible has been previously explored~\cite{ harrington2004:phd, dennis2005:phd, duclos2010fast},
but in order to make such a decoder fast one needs to be able rapidly 
update the clusters as they grow. This dynamical update gives the dominant contribution to the decoding complexity.

\medskip
To define how we can identify invalid clusters of erasures, we state our first lemma.

\begin{lemma}[even vs odd clusters] \label{lemma:erasure_decoding_parity_condition}
Let $\e$ be a connected subset of edges and let $\sigma$ be a set of syndrome 
vertices included in $\e$.
There exists a  $Z$-error $E_Z \subset \e$ of syndrome $\sigma$ if and only if
the cardinality of $\sigma$ is even.
\end{lemma}

This is straightforward to prove. 
We define a cluster to be a connected component of erased qubits in the subgraph $(V,\e)$.
These clusters can be either a connected subgraph induced by a subset of erased edges, 
or an isolated vertex.
If no Pauli error is present, it must be the case that the cluster supports an even number
of syndrome bits. If the cluster supports an odd number of syndrome vertices then we identify it as `invalid', at least 
one error chain must terminate in this cluster. The erasure decoder cannot be applied
directly to odd clusters

\medskip
Since we cannot apply the erasure decoder to odd clusters directly, instead we grow these clusters
by adding edges to the erasure until they connect with another odd cluster. When two odd clusters merge
the resulting cluster is even, and can therefore be corrected. 
Note that a single vertex outside the erasure supporting a syndrome bit is an odd cluster.

\begin{figure}
\includegraphics[width=\columnwidth]{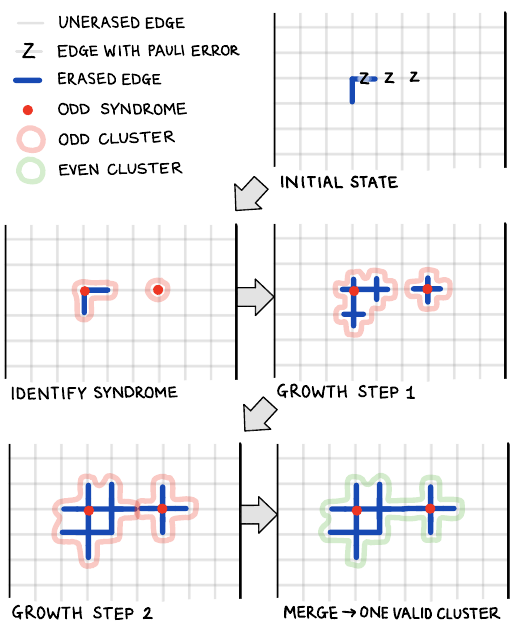}
\caption{\label{fig:schematic} Schematic representation of syndrome validation.
}
\end{figure}

\medskip
Algorithm~\ref{algo:uf_naive} describes this procedure. Steps 1-7 
perform syndrome validation, and in step 8 the erasure decoder is applied. 
Lemma~\ref{lemma:erasure_decoding_parity_condition} proves that we can apply the erasure decoding at the end of Algorithm~\ref{algo:uf_naive} 
and that this final step will return a correction $\mathcal{C}$ that is consistent with the syndrome.
Figure~\ref{fig:schematic} shows
an example of the growing and merging procedure. 

\medskip
We will later discuss the detail of the implementation required to achieve a low complexity, but first we
focus on the error tolerance of the algorithm. 

\begin{algorithm}[h]
  \SetAlgoLined
  \DontPrintSemicolon

  \SetKwInOut{Input}{input}\SetKwInOut{Output}{output}
  \Input{The set of erased positions $\e \subset E$ and the syndrome $\sigma \subset V$ of an error $E_Z$.}
  \Output{An estimation ${\cal C}$ of $E_Z$ up to a stabilizer.}
  \BlankLine
	
  Create the list of all odd clusters $C_1, \dots, C_m$, and initialize the modified erasure $\e'=\e$. \;
  \While{there exists an odd cluster}{
    \For{all odd cluster $C_i$}{
      Grow $C_i$ by increasing its radius by one half-edge. \;
      If $C_i$ meets another cluster, fuse and update parity. \;
      If $C_i$ is even, remove it from the odd cluster list. \;
  	}
  }
  Add full edges that are in the grown clusters to $\e'$. \;
  Apply the peeling decoder to the erasure to find $\mathcal{C}$. \;
  
	\caption{Union-Find decoder -- Naive version}
	\label{algo:uf_naive}
\end{algorithm}

\section{Decoder Performance\label{sec:performance}}

We now discuss the performance of the decoder in the low error regime, and 
show that it performs just as well as the most likely error (MLE) decoder below the minimum distance.
MLE decoder returns the most likely error configuration given the observed syndrome. 
The MWPM-decoder is an implementation of an MLE decoder.

\medskip
For any code, the MLE decoder can correct any error configuration, $E_Z$, with a weight
of up to $(d-1)/2$, where $d$ is the minimum distance of the code.
For the erasure channel, any erasure pattern
of up to $d-1$ qubits can be corrected. 
Moreover, both of these bounds are tight. Reaching these values 
with an efficient decoding algorithm is evidence of good performance.

\medskip
In our mixed noise model, MLE decoder can correct any combination
of $t$ erased qubits and $s$ $Z$-errors (outside the erased set) as long as
$t + 2s<d$. This bound, generalizing both previous cases, is also tight.

\begin{theo} \label{theo:UF_perf}
If $t + 2s<d$, Algorithm~\ref{algo:uf_naive} can correct any 
combination of $t$ erased qubits and $s$ $Z$-error.
\end{theo}

This algorithm performs well for both Pauli errors, erasures and combinations of the two. 

\begin{proof}
For the case of erasures the proof is straightforward since the erasure decoder can be called 
immediately without going through steps 1. to 7.
Consider next the other extreme where only Pauli errors occur and denote the error by
$E_Z$.
A cluster $C$ grows in Algorithm~\ref{algo:uf_naive} (Step 4.), 
when there are an odd number of syndrome vertices contained within it.
This implies that there exists at least one path in the support of $E_Z$
connecting a vertex of $C$ to a vertex outside of $C$.
Therefore, when a cluster grows, at least one new half-edge of $E_Z$ is covered.
After at most $2s$ rounds of growth, the grown cluster covers 
the entire error $E_Z$ and there can be no more odd clusters left to grow.
At the end of the growing procedure, the diameter of the largest erased
cluster is at most $2s$ edges ($4s$ half-edges). 
By erasing this cluster, we obtain an erasure pattern that covers $E_Z$ 
and that does not cover a non-trivial logical error since $2s < d$. 
When the peeling decoder is run in the final step it must therefore succeed at identifying $E_Z$ up 
to a stabilizer. This argument relies on the optimality of the peeling decoder proven in \cite{delfosse2017:linear_decoding}

The general argument for a combination of $s$ errors and $t$ erasures is similar. 
Growing the clusters increases the diameter of the largest cluster by at 
most $2s$. It is then upper bounded by $2s+t<d$. Just as in the case of only Pauli errors, the final 
step returns an error equivalent to $E_Z$, up to a stabilizer.
\end{proof}

\section{Achieving almost-linear complexity\label{sec:complexity}}

Our next goal is to show that our decoder can be implemented in almost-linear time
in the number of qubits, $n$, by exploiting Union-Find data-structure 
algorithms~\cite{galler1964:union_find_algorithm, tarjan1975:union_find_complexity}.

\subsection{Union-Find algorithm for cluster growth}
The key function of the decoder is to grow clusters, and fuse them when they
meet. For this we need two things: a function $\Union(u, v)$ that performs the 
fusion operation on two clusters
$C_u$ and $C_v$, and a function $\Find(v)$ that identifies the cluster to which
vertex $v$ belongs. The $\Find()$ function
allows us to distinguish clusters from one another, since we only wish to fuse
clusters when they are distinct.  
When an edge, $(u,v)$ is added to a growing cluster, we call the function
$\Find(u)$ and $\Find(v)$ on its two endpoints. 
If the endpoints belong to the same cluster, $\Find(u)=\Find(v)$,
in which case we do nothing. If $\Find(u)\neq\Find(v)$, they belong to different clusters and we must fuse
$C_u$ and $C_v$. The complexity of this subroutine provides the leading order in the complexity of 
our decoder.

\medskip
{\bf A naive algorithm:}
We first describe a naive implementation of the $\Union()$ and $\Find()$ functions.
Let us store an index $\Find(v)$ for each vertex $v$
in a look-up table of size $|V|$.
This makes the cluster identification trivial.
However, when $\Union(u, v)$ is called one must update 
the cluster indices $\Find(w)$ for all the vertices $w$ of one the 
two clusters $C_u$ or $C_v$ in order to correctly update the state. 
This might require a number of updates of cluster indices $\Find(v)$,
which is itself linear in $n$. 
In the worst case, we call $\Union$ up to $n-1$ times, yielding a 
quadratic overall complexity, $O(n^2)$.

\subsection{Implementation}
We now describe the data structure, and steps required in the implementation
to achieve an almost-linear complexity. 

\medskip
{\bf Weighted Union:}
In order to reduce the complexity due to updating the cluster index
after merging, we can choose to always select the smallest component of the two 
clusters, $C_u$ and $C_v$, to update.
The size of the small cluster at least doubles at each call of $\Union()$,
which means that every vertex index is updated at most $O(\log(n))$ times,
reducing the complexity of the Union-Find update to $O(n \log n)$.
To do this it is also necessary to store the size of each component, but this
does not affect the complexity, as we can simply add them to our look-up table. 

\begin{figure}
\includegraphics[width=\columnwidth]{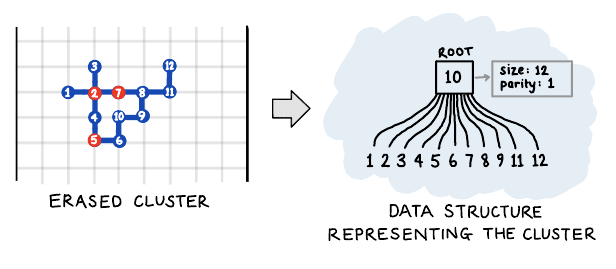}
\caption{\label{fig:data_structure} {\bf A cluster-tree data structure.}  The tree is initialized with depth one but 
this may increase during the decoding procedure.
The size and parity are stored at the root of the cluster-tree.}
\end{figure}

\medskip
{\bf Tree representation:}
We now consider how the clusters can be stored in memory in order
to speed up the index update in $\Union()$. The representation we choose
increases the cost of the $\Find()$ function, but overall the complexity 
is reduced. 
We represent each cluster $C$ as a tree that we call a {\em cluster tree}. 
The vertices of the cluster-tree correspond to the vertices of the
cluster $C$ of the lattice, however, the cluster-tree is an arbitrary 
tree and does not have to respect the lattice structure. An example of this
data structure is shown in Figure~\ref{fig:data_structure}. 
An arbitrary vertex $u$ of $C$ is chosen as the {\em root} of the 
cluster-tree and is used to identify the cluster. The size of the cluster and its parity are also stored at its root.
$\Find(v)$ returns the root of $u$, for which we must traverse the tree from $u$ to its root.
The cost of $\Find()$ is therefore given by the depth of the tree. 
To minimize this cost, the cluster-trees are initialized with depth 1 at the beginning of the 
Union-Find decoder. 
When $\Union(u, v)$ is applied, one must first check whether $u$ and $v$ belong
to the same cluster. To do this we call $\Find(u)$ and $\Find(v)$ and walk through the trees 
from $u$ and $v$ to their respective roots. 
If these components are distinct, the trees are merged in two steps. 
First, the smallest cluster-tree (say $C_v$) is added as a subtree of the root 
$\Find(u)$ of the largest cluster-tree ($C_u$).
Then, the size and parity stored at the root is updated. Path compression
can then be applied to minimize the depth of the new clusters.

\medskip
{\bf Path compression:}
When $Find(u)$ is called, we step through the tree until we reach its root. 
Once the root of $u$ is reached we compress the by pointing all the vertices 
encountered along the path directly to the root. 
This only doubles the cost $Find()$ but speeds up any future call to the function that passes
through any part of the compressed path. Using path compression alone, without
weighted union also leads to a complexity of $O(n \log n)$. However, when combining both
procedures together there is a further reduction of the complexity. 
The analysis of Union-Find algorithms with both weighted union and 
path compression is quite involved, Tarjan proved that remarkably the worst-case 
complexity is $O(n \alpha(n))$ 
\cite{tarjan1975:union_find_complexity}. It has also been shown that this result is optimal \cite{tarjan1979:union_find_lower_bound, fredman1989:union_find_lower_bound}.

\medskip
{\bf Support of the growing clusters:}
The cluster-trees encode the set of vertices of the clusters.
The edges are represented by a look-up table, $\Support$, of size $|E|$
that stores the state of an edge, which can take one of 
three values: {\em Unoccupied}, {\em Grown} or {\em Half-grown}.
When an edge is half-grown also store the vertex of the lattice
from which it was grown.
$\Support$ is initialized with all erased edges marked as {\em Grown}
and all other edges being {\em Unoccupied}. As the algorithm proceeds, 
half-edges are grown, growing by one half-edge twice produces a grown edge.

\medskip
{\bf Boundary representation:}
In order to preserve the low complexity of the dynamical fusion 
using Union-Find, we must be careful when updating $\Support$. In each round
of growth we must identify the boundary vertices of a cluster from which we will
grow new edges. By a {\em boundary vertex}, we mean a vertex of the lattice such that at least one of the
incident edges is not in the erasure, $\e$. One way to do this would be to explore the cluster to identify
the boundary, but this would induce a linear cost in the size of the cluster 
in each round, resulting in a quadratic complexity overall.
We now describe a more efficient strategy.

\medskip
We can avoid recomputing the boundary of the cluster by
storing a list of boundary vertices for each root. 
To grow a cluster, we must then simply iterate over this list and grow the incident edges.
When clusters are fused, it is necessary to update the boundary lists. We now describe a 
subroutine $\Grow(\cal L)$ that can achieve this in linear time. 
This function takes a list ${\cal L} = \{u_1, u_2, \dots, u_m\}$ of cluster roots as an input, 
each with its associated boundary list. The function $\Grow(\cal L)$ increases the radius of 
these clusters by one half-edge, fuses the clusters that meet and 
updates the clusters and their boundary lists.
The function runs in five steps.
{\begin{itemize}
\item 
{\em $(i)$ Growth:} Grow all the clusters of root $u_i$ by one half-edge. This is done by running over the
lists of boundary vertices and by growing the incident half-edges in the table $\Support$. 
This step returns a list of all the newly grown edges that connect two distinct clusters. 
Call these edges {\em fusion edges}.
\item 
{\em $(ii)$ Fusion of clusters:} Run over fusion edges 
$e=\{u,v\}$ and if $u$ and $v$ belong to distinct clusters then merge them with $\Union(u,v)$, otherwise remove $e$ from the list of fusion edges.
\item 
{\em $(iii)$ Fusion of boundary lists:}
Run over fusion edges (updated in (ii)) and for each $e=\{u,v\}$, 
append the boundary list of the smallest cluster $C_u$ or $C_v$ (before fusion) 
at the end of the boundary list of the largest one (before fusion).
\item
{\em $(iv)$ Update roots:} Replace each element $u_i$ of the root list $\cal L$ by $\Find(u_i)$.
\item 
{\em $(v)$ Update Boundary lists:} Run once more over each boundary list and
remove the vertices that are no longer boundary vertices.
\end{itemize}

We remark that new boundary vertices are added when merging the boundary list with the
boundary list of a neighbor cluster. We only need to remove vertices from those lists
in the last step to update them.

\medskip
{\bf Avoiding growth duplication:}
In order to simplify the description of the algorithm, we have so far omitted a detail of the implementation.
If two clusters of the list are fused during step {\em (ii)}, then updating their roots 
in {\em (iv)} results in a duplicated root in the list.
We can avoid this issue by also storing in a look-up table an indicator that marks 
the vertices of the odd-root list at the beginning of each round of growth. 
Maintaining this extra look-up table (of linear-size) does not increase the complexity of the algorithm. 
This modification allows us to detect the presence of $Find(u)$ in 
the list before replacing $u$ by $Find(u)$. If it is already present, we simply remove $u$
from the list, and avoid creating a duplicate.

\subsection{Summary of data structure and algorithm}

Each cluster $C$ in the lattice is encoded using the following data:
\begin{itemize}
\item {\bf Cluster-tree:} A tree whose vertices encode the vertices of $C$ with an arbitrary root.
\item {\bf Size and Parity:} The size of the cluster and the parity of $C$ are stored at the root of the cluster-tree.
\item {\bf Support:} A look-up table that stores the state of each edge in the growing clusters.
\item {\bf Boundary List:} The list of all the boundary vertices of $C$.
\end{itemize}

A full version of the algorithm we have described is given in Algorithm~\ref{algo:uf_linear}. Let us now summarize the contribution of each part of the algorithm to the overall complexity.

\medskip
{\bf Complexity of the full algorithm: } Algorithm~\ref{algo:uf_linear} 
can be decomposed into three blocks. 
The first block contains lines 1-2 and initializes the clusters.
Block 2 contains lines 3-11 and relies on a Union-Find algorithm that we have 
just described. The complexity of this subroutine provides the leading order in the complexity of 
our decoder. In the third block, line 12, the erasure decoder is applied. 

\medskip
Creating the list of clusters can be achieved in linear time by exploring the connected
components of the sublattice of erased edges \cite{hopcroft1973:graph_components_algo}.
During this same exploration, we can compute the size and syndrome parity of the cluster, and store it at 
the root, which does not affect the complexity. 
The table $\Support$ is also initialized in linear time.
The second block grows and merges odd clusters until they disappear and 
runs from line 3 to 11. We have discussed how the Union-Find algorithm can be used
to achieve this in $O(n \alpha(n))$ time. 
Each of the $n$ vertices can be at a boundary during at most two rounds of growth, and during 
the growing procedure each list is iterated over $O(1)$ times. 
The complexity of growing $\Support$ and updating the boundary
lists is therefore $O(n)$. 
The support of the clusters and the boundary lists can be updated in linear time $O(n)$
using $\Grow()$. 
The last two instructions are simply the erasure decoder and are already known 
to have linear complexity \cite{delfosse2017:linear_decoding}.
The dominant contribution to the complexity is therefore the $O(n \alpha(n))$ 
due to the Union-Find algorithm used to keep track of the cluster-trees during syndrome validation.

\begin{algorithm}[h]
  \SetAlgoLined
  \DontPrintSemicolon

  \SetKwInOut{Input}{input}\SetKwInOut{Output}{output}
  \Input{The set of erased positions $\e \subset E$ and the syndrome $\sigma \subset V$ of an error $E_Z$.}
  \Output{An estimation ${\cal C}$ of $E_Z$ up to a stabilizer.}
  \BlankLine

  Initialize cluster-trees, $\Support$ and boundary lists for all clusters. \;
  Create the lists ${\cal L}$ of roots of odd clusters. \;
  \While{${\cal L}$ is not empty}{
    (o) Initialize the fusion list $\cal F$ as an empty list. \;
    (i) For all $u \in {\cal L}$, grow the cluster $C_u$ of a half-edge in the Table $\Support$. If a new grown edge $e$ is added in $\Support$ then add $e$ to the fusion list $\cal F$. \;
    (ii) For all $e=\{u,v\} \in {\cal F}$, if $Find(u) \neq Find(v)$ then applies $\Union(u,v)$ to merge the cluster $C_u$ and $C_v$. If $Find(u) = Find(v)$ then remove $e$ from the list $\cal F$. \;
    (iii) For all $e=\{u,v\} \in {\cal F}$, read the sizes of the clusters $C_u$ and $C_v$ stored at the roots (uses $Find()$) and append the boundary list of the smallest cluster at the end of the boundary list of the largest one. \;
    (iv) Replace each root $u \in {\cal L}$ by $\Find(u_i)$. (in a way that does not create duplicated elements) \;
    (v) For all $u \in {\cal L}$, remove the vertices of the boundary list of $u$ that are not boundary vertices. \;
    (vi) For all $u \in {\cal L}$, if $C_u$ is an even cluster then remove it from $\cal L$. \;
  }
  Erase all the edges that are fully grown in $Support$. \;
  Apply the peeling decoder to the erasure. \;

  \caption{Union-Find decoder -- Almost-linear time version}
  \label{algo:uf_linear}
\end{algorithm}

\section{ Weighted growth version of the Union-Find decoder\label{sec:weighted_growth}}
The algorithm we have described so far grows all (odd) clusters uniformly in each round.
We now discuss one simple way in which this growth strategy can be altered 
to improve the performance of the decoder.
We recall that our basic argument in the proof of Theorem~\ref{theo:UF_perf}
was that if a cluster is odd there exists at least one path of errors connecting this cluster to
a vertex outside the cluster. When the error rate is small, it is likely
that only one error chain terminates in the cluster. When the cluster grows, we add some number
of edges that is proportional to the number of boundary vertices. If we add $b$ 
new edges, then only $1/b$ of those edges correctly cover the error. The larger
the boundary of the cluster, the more 'incorrect' edges we are adding. 
By growing smaller clusters first, fewer erasures will be added in total, which increases
the chance success in the the final erasure decoding step .

\medskip
The {\em Weighted Growth} version of the Union-Find decoder is achieved by 
always growing the cluster with the smallest boundary size. 
This small modification improves the threshold of the 2d toric code from $9.2\%$ to $9.9\%$
for phase-flip errors and from $2.4\%$ to $2.6\%$ with faulty-syndrome measurements.

\medskip
The Weighted Growth Union-Find decoder can also be implemented in almost-linear time.
To achieve this complexity, we need to sort the clusters by boundary size. This can be done in linear time in $n$ because clusters boundary size are integer with values between 1 and $O(n)$.

\section{Application to Quantum Computing\label{sec:application}}

We have so far described the decoder only for the 2d surface code. To be of use for quantum computing applications, where measurements may be faulty, we must be able to solve the 
(2+1)-dimensional variant of the decoding problem \cite{dennis2002topological}. Multiple rounds of syndrome measurement on the surface code produce a three-dimensional cubic lattice of syndrome outcomes, where the x-y directions correspond to the physical code, and the third dimension represents time. Space-like edges in this syndrome lattice correspond to physical errors on the qubits of the code, while time-like edges correspond to measurement errors. The decoder as we have described it requires no adaption to be applied to the 3d lattice. 

\medskip
Conceptually we can understand the decoder as adding virtual erasures to the syndrome information that include the support of the error operator, and we can extend this interpretation to the 3d case. Erasing the space-like edges of the syndrome graph corresponds to erasure of a physical qubit, while erasing a time-like edge of the syndrome graph corresponds to the erasure of a measurement outcome - a classical erasure. Physically such a measurement erasure corresponds to the case that no stabilizer outcome was recorded.

\begin{figure}
\includegraphics[width=\columnwidth]{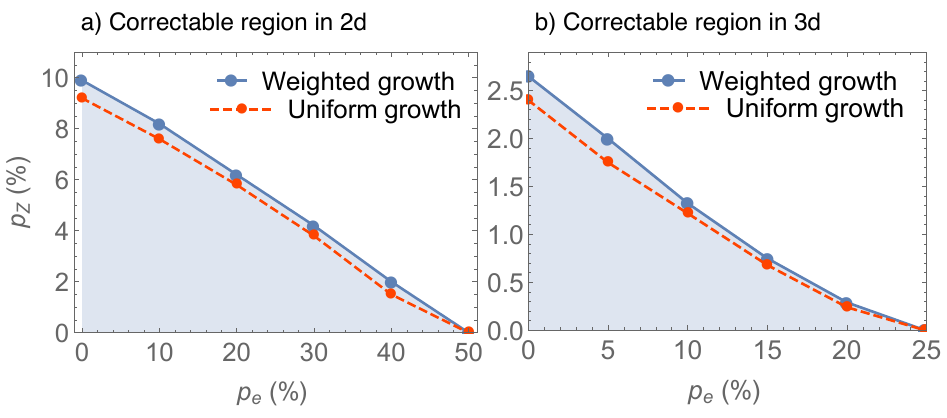}
\caption{\label{fig:plots} {\bf Performance results.} a) Results of numerical threshold simulations on the 2d toric code under an error model of erasure and Pauli error at independent error rates $p_e$ and $p_Z$. The shaded area indicates the correctable region, which is below threshold. Example threshold plots can be seen in the Appendix. b) Results of numerical threshold simulations of the 2+1d toric code under an error model of erasure and phenomenological error at independent error rates $p_e$ and $p_Z$. 
}
\end{figure}

\medskip
We performed numerical simulations to quantify the performance of the decoder in both 2d and 3d. In 2d each qubit is treated as erased with a probability $p_e$ and acquires a Pauli $Z$ error with a probability $p_Z$. In 3d we consider making $L$ rounds of repeated syndrome measurements, for a lattice of dimension $L$. Each measurement outcome is erased with probability $p_e$ and returns the incorrect value with probability $p_Z$. Between rounds of measurement each qubit is erased with probability $p_e$ and acquires a Pauli $Z$ error with probability $p_Z$. This corresponds to the phenomenologial error model in the case that $p_e=0$. More details of the numerics are described in Appendix~\ref{sec:numerics}. The results of threshold calculations are shown in Figure~\ref{fig:plots}a) and b), defining a correctable region in the space of erasure and Pauli error. In 2d, with no erasure, we find $p_{th} = 9.9\%$ and in 3d we find $p_{th} = 2.6\%$. This is only a small reduction in the decoder performance compared with MWPM (see Table-\ref{tab:thresholds}). Below threshold the logical error rate is well behaved, showing a clear exponential supression with increasing lattice size.

\medskip
We measured the average running time to perform decoding as a function of the number of qubits, which gives numerical evidence of a linear average complexity. Figure~\ref{fig:time_scaling} shows the running time, for a C implementation of the decoder running on a single 2.9 GHz Intel Core i5 CPU.

\subsection{Application beyond the surface code}

The functionality of the Union-Find decoder extends beyond Pauli errors, as the decoder naturally handles a mixed noise model of erasure and Pauli error. Barrett and Stace \cite{stace2010:loss_long, barrett2010:loss} first introduced an algorithm to decode the Raussendorf lattice \cite{raussenfort2007:cluster, raussenfort2006:ft, raussenfort2007:ft} in the presence of erasure, which combines MWPM with geometric deformations of the lattice around erased qubits, and runs in $O(n^3)$ time. This noise model is particularly relevant for photonic measurement based quantum computing 
 \cite{knill2001:LOQC, nielsen2004:LOCC_cluster, browne2005:LOCC_efficient, kieling2007:LOCC_percolation}, where qubit erasure is likely to be a dominant source of error. However, in these architectures the complexity of the Barrett-Stace algorithm would make it challenging to implement fast enough to keep up with the natural clock-speed of a quantum computer. The Union-Find decoder provides a significant decrease in computational requirements for decoding in a photonic architecture, with only a small decrease in threshold. 

\medskip
As well as its speed, one advantage of the Union-Find decoder is its immediate flexibility to other geometries and dimensions. The Union-Find decoder requires only an underlying graph structure, and therefore requires no alteration to run in any geometry as long as the errors form string-like objects. This makes it a simple tool for use in complex geometries, such as hyperbolic codes~\cite{freedman2002z,zemor2009cayley,delfosse2013tradeoffs,breuckmann2016:hyperbolic,breuckmann2017hyperbolic}. In contrast, MWPM must in general be combined with costly path finding algorithms to compute the distance between syndrome measurements. Here we have only simulated the performance on a torus, but non periodic boundary conditions can be handled by allowing clusters to become valid by merging with a rough boundary. Furthermore, the decoder can also be used on a syndrome graph of any dimension, although we expect that the threshold performance of the Union-Find algorithm, at least in its basic form, will decrease relative to perfect matching as the dimension of the space increases. 

\medskip
Beyond surface codes, the Union-Find decoder can also be used to decode the 2D color code, when used in combination with the method introduced in~\cite{delfosse2014decoding} for projecting color codes onto surface codes. This procedure does not affect the complexity of the algorithm. By numerically simulating decoding on a hexagonal lattice, we find a threshold for the [6,6,6] color code of 8.4\% under i.i.d Pauli $Z$ error.

\section{Conclusion}

We have presented a decoder with a high threshold which has a considerably lower worst-case complexity than any other existing practical decoder. 
Indeed, we are very close to the the best possible complexity for any decoder that is not parallelized. 
Since one must at a minimum iterate at least once over the syndrome, the lowest possible time is O(n). 
But we are interested in practical applications, and complexity is not the only figure of merit. It is 
important that our algorithm also has a small constant overhead, making it fast in practice as well as in theory.

\medskip
Instead of designing a fast decoder which approximates the MWPM decoder as we did in this work, one could consider a parallel MWPM decoder~\cite{fowler2015:parallel_MWPM} which achieves a linear average-case complexity. It would be interesting to investigate the parallelization of the Union-Find decoder.

\medskip
Many questions remain to be answered. Our implementation achieved linear complexity, but was not heavily optimized for speed. Work should be done to understand how much this can be improved, in particular by implementing the algorithm in hardware. In terms of error tolerance, we have studied only simple noise models, and the performance under circuit-level error will need to be studied to draw a more meaningful comparison with the thresholds of other decoders. 

\medskip
There are many challenges still to solve before an error corrected quantum computer can be realized, and one of these is decoding. Our algorithm has important practical implications in achieving this goal, and is a significant step towards overcoming the hurdle of fast decoding in real quantum devices. 

\medskip
{\em Acknowledgement -- The authors would like to thank 
Eric Johnson and Chris Dawson for valuable discussions, and 
Terry Rudolph for first introducing them to the question of 
error correction in photonic devices. The authors would like to 
thank Aleksander Kubica for his comments on a preliminary 
version of this paper.
ND acknowledges funding provided by the Institute for Quantum Information and Matter, an NSF Physics Frontiers Center (NSF Grant PHY-1125565) with support of the
Gordon and Betty Moore Foundation (GBMF-2644).}


\newcommand{\SortNoop}[1]{}

\begin{appendix}

\section{Ackermann's function}

Different versions of Ackermann's original function exist, though they generally only different 
in a constant. 
In this section, we describe the version used by Tarjan to obtain an optimal 
upper bound on the complexity of Union-Find algorithms \cite{tarjan1975:union_find_complexity}.
Ackermann's function is a two-parameter function defined for any pair 
$(i, j)$ of non-negative integers by the following relations,
\begin{align*}
& A(0, i) = 2i\\
& A(i, 0) = 0\\
& A(i, 1) = 2\\
& A(i, j) = A(i-1, A(i, j-1))
\end{align*}
From this recursive definition, we find, for instance, that for all $j \geq 1$
$$
A(1, j) = A(0, A(1, j-1)) = 2A(1,j-1) = 2^j \cdot
$$
A similar calculation shows that $A(2,1) = 2, A(2,2) = 4$ and
$$
A(2, j) = 2^{{{2^{2}}^{-}}^{2}} \text{ with } j \text{ twos} \cdot
$$

We call the inverse of Ackermann's function, which we denote $\alpha(n)$, the
value 
$$
\alpha(n) = \min \{ i | \ | A(i, 4) \geq \log_2 n \}
$$
Ackermann's function grows amazingly quickly. The first terms to appear in the
definition of $\alpha$ are
$$
A(1,4) = 2^4,
$$
and
$$
A(2,4) = 2^{2^{2^{2}}} = 2^{16} = 65536.
$$
Then, we obtain $A(3,4) = A(2, A(3,3)) = A(2, A(2, A(3,2))) = A(2, A(2,4))$
and using $A(3,2) = 4$, we find
$$
A(3,4) = 2^{{{2^{2}}^{-}}^{2}} \text{ with 65536 twos} \cdot
$$
This number is so large that we never hit $A(3,4)$
in any practical situation, making $\alpha(n) \leq 3$.

\section{Numerical results \label{sec:numerics}}

\begin{figure}
\centering
\includegraphics[width=0.44\columnwidth]{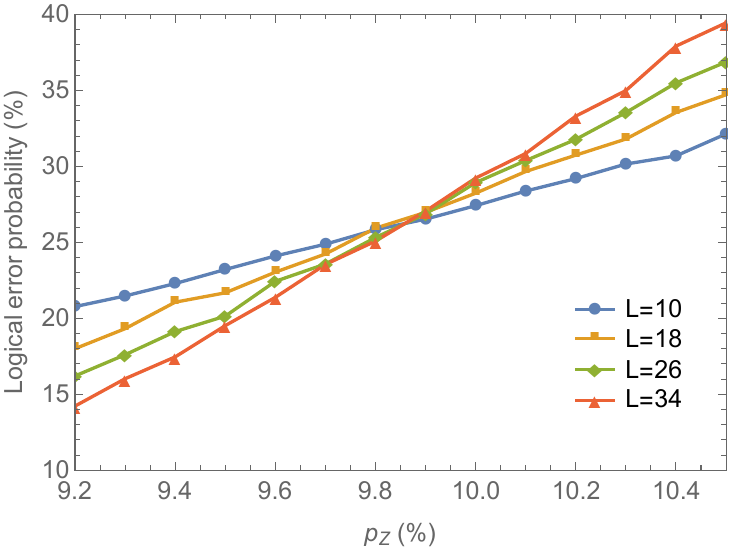}
\includegraphics[width=0.45\columnwidth]{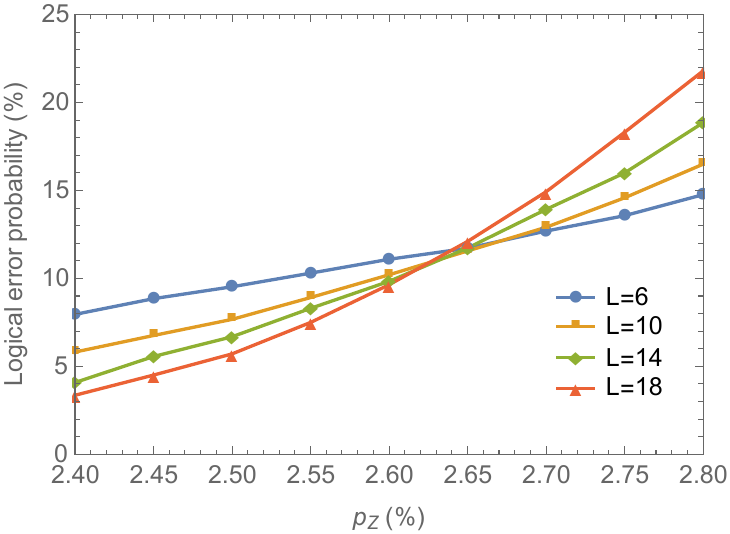}

(a) \hspace{0.3\columnwidth} (b)

\caption{
(a) Threshold of the 2d toric code for $Z$-type error with no erasure.
(b) Threshold of the 3d toric code for $Z$-type error with no erasure.
}
\label{fig:threshold0}
\end{figure}

\begin{figure}
\centering
\includegraphics[width=0.75\columnwidth]{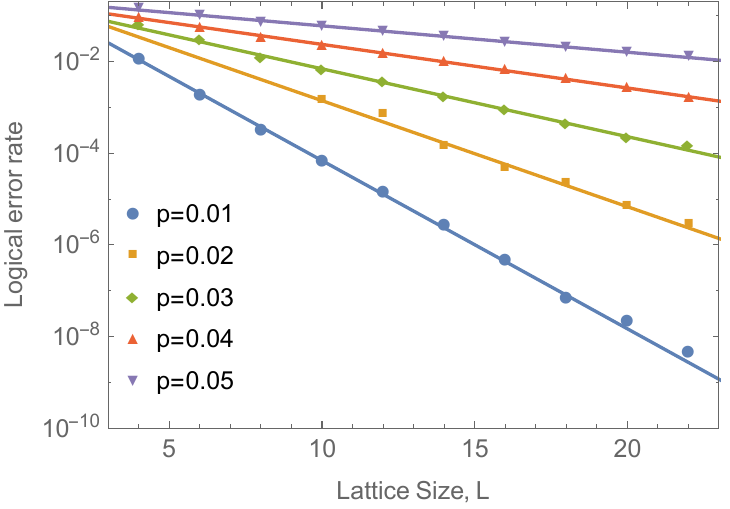}
\caption{\label{fig:scaling} Below threshold performance for the 2d toric code with $p_e=0.1$. Below the threshold the logical error rate is exponentially suppressed as the lattice size increases. We perform up to $10^{9}$ Montecarlo trials per data point. }
\end{figure}

Here we show an example of the threshold plots for the numerics that were used to generate the data shown in Figure~\ref{fig:plots}. All data points were computed by repeated montecarlo simulations of random erasure and noise on a surface code followed by decoding. The simulations were repeated until at least 10,000 failures had been observed. Figure~\ref{fig:threshold0}(a) shows the threshold for the 2d toric code, when $p_e=0$, for which we find a threshold of $9.9\%$. Figure~\ref{fig:threshold0}(b) shows the threshold for the 2+1d toric code, when $p_e=0$, for which we find a threshold of $2.6\%$. 

Figure~\ref{fig:scaling} shows the below threshold scaling of the where $p_e=0.1$. We see an exponential suppression in the logical error rate with increasing lattice size, strong evidence of the threshold behavior.

\end{appendix}

\end{document}